\documentclass[aps,prl,twocolumn,groupedaddress,amsmath,amssymb,superscriptaddress,showpacs]{revtex4}
\usepackage{eufrak}
\usepackage{graphicx}
\usepackage{dcolumn}
\usepackage{bm}
\usepackage{multirow}
\usepackage{mathrsfs}
\usepackage{calrsfs}
\usepackage[T1]{fontenc}
\usepackage{color}
\usepackage{epsfig}
\begin{document}

\title{Enhanced diffusion and ordering of  self-propelled rods}
\author{Aparna Baskaran}
\affiliation{Physics Department, Syracuse University, Syracuse NY
13244}
\author{M. Cristina Marchetti}
\affiliation{Physics Department and Syracuse Biomaterials
Institute, Syracuse University, Syracuse, NY 13244, USA}

\date{\today}

\begin{abstract}
Starting from a minimal physical model of self propelled hard rods
on a substrate in two dimensions, we derive a modified
Smoluchowski equation for the system. Self -propulsion enhances
longitudinal diffusion and modifies the mean field excluded volume
interaction. From the Smoluchowski equation we obtain hydrodynamic
equations for rod concentration, polarization and nematic order
parameter.  New results at large scales are a lowering of the
density of the isotropic-nematic transition and a strong
enhancement of boundary effects in confined self-propelled
systems.
\end{abstract}


\maketitle

Self propelled particles consume energy from internal or external
sources and dissipate it by actively moving through the medium
that they inhabit. Assemblies of interacting self-propelled
particles (SPP) exhibit rich collective behavior, such as
nonequilibrium phase transitions between disordered and ordered
(possibly moving) states and novel long-range correlations.
Biologically relevant systems that belong to this class include
fish schools, bird flocks \cite{TonerRev}, bacterial colonies
\cite {bactexpts} and cell extracts of cytoskeletal filaments and
associated motor proteins \cite{actinexpts}. A non-living
realization may be a vibrated monolayer of granular rods
\cite{rodsexpt}. Collections of SPP have been the focus of
extensive experimental \cite{actinexpts,rodsexpt,experiments} and
theoretical studies in recent years. A number of distinct
theoretical approaches have proved fruitful for understanding the
complex dynamics of these nonequilibrium systems. These include
numerical studies of simple models
\cite{Orsogna,Chate2, Lipowsky, Peruani06}, inspired by the
seminal work of Vicsek \cite{Vicsek}, and phenomenological
continuum theories based on general symmetry arguments
\cite{RamaswamyPRL}. Recent work on deriving the hydrodynamic
equations from specific microscopic models has led to some insight
into the origin of the collective behavior of these systems \cite
{Marchetti,Aranson1, Boltzmann, Aparna2, Shelley}. An important
open question that we address here is the interplay between self propulsion and steric
effects arising from the shape of the particle in controlling the
large scale physics.

In this paper we consider a physical model of self-propelled hard
rods that interact with each other solely through excluded volume.
 The rods move on a passive substrate.
Self-propulsion is modeled as a nonequilibrium velocity $v_0$
along the direction of the rods' long axes. The goal of our work
is to understand how self-propulsion modifies the diffusion
processes and the mean-field Onsager excluded volume
interaction~\cite{EdDoiBook}. Using the tools of nonequilibrium
statistical mechanics we derive a \emph{modified} Smoluchowski
equation that differs from the familiar version for thermal hard
rods \cite{EdDoiBook} in three respects. The first and obvious
modification is a convective mass flux at the self-propulsion 
speed $v_{0}$ along the direction of orientation of the rod.
Secondly, self-propulsion enhances the longitudinal diffusion
constant $D_{\parallel }$ of the rods, according to $D_{\parallel
}\rightarrow D_{\parallel }(1+v_{0}^{2}/k_{B}T)$. This
enhancement arises because self-propelled particles perform a
\emph{persistent random walk}, as recently pointed out by other
authors \cite{PersitentRW}. Finally,  the momentum exchanged by
two rods upon collision is rendered highly anisotropic by
self-propulsion thus modifying the Onsager form of the excluded
volume interaction. This leads to novel anisotropic forces and
torques from steric repulsion in the Smoluchowski equation.

These modifications of the Smoluchowski equation have dramatic
consequences for the properties of the system on hydrodynamic
scales. This is illustrated by two examples.  First, we show that
the additional momentum transfer from self-propulsion lowers the
density of the isotropic-nematic transition, thereby providing a
microscopic identification for the physical mechanism responsible
for the enhancement of orientational order observed in numerical
simulations of motility assays \cite{Lipowsky}. Secondly, we
demonstrate that self-propulsion greatly enhances the effect of
confinement and the role of boundaries.

\paragraph{The microscopic model.}

\begin{figure}[tbp]
\centerline{\epsfxsize=8cm \epsfbox{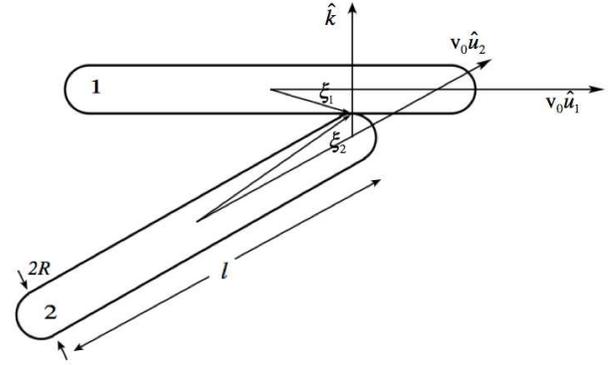}} 
\caption{ A cap-to-side collision of two self-propelled hard rods
(the width of the rod is exaggerated for clarity). $\widehat{\bf k}$ is a
unit vector from rod 2 to rod 1  normal to the point
of contact. Points on the side of the rods are identified by
vectors ${\bm\protect\xi}_i$. } \label{Fig:collision}
\end{figure}

We consider quasi two-dimensional hard rods of length $\ell $
and thickness $2R$ confined to a plane,
as shown in Fig.~\ref{Fig:collision}. The $i$-th
rod is characterized by the
position $\mathbf{r}_{i}$ of its center of mass and a unit vector $\mathbf{%
\hat{u}}_{i}=\left( \cos \theta _{i},\sin \theta _{i}\right) $
directed along its long axis. Each rod free-streams on the
substrate, until it collides with another rod. The collision
results in instantaneous linear and angular momentum transfer such
that the total energy, linear and angular momenta of the two rods
are conserved. The microdynamics of the system is governed by
coupled Langevin equations,
\begin{eqnarray}
&&\frac{\partial \mathbf{v}_{i}}{\partial t}=-\sum_{j}T\left(
i,j\right)
\mathbf{v}_{i}+F\widehat{\mathbf{u}}_{i}-\bm\zeta _{i}\cdot \mathbf{v}_{i}+%
\bm\eta _{i}\left( t\right) ,  \label{2} \\
&&\frac{\partial \omega _{i}}{\partial t}=-\sum_{j}T\left(
i,j\right) \omega _{i}-\zeta ^{R}\omega _{i}+\eta _{i}^{R}\left(
t\right) \;,  \label{3}
\end{eqnarray}
where $\mathbf{v}_{i}=\partial _{t}\mathbf{r}_{i}$ and $\omega
_{i}=\partial _{t}\theta _{i}$ are the center of mass and angular
velocities, $\bm\zeta _{i}$ is the friction tensor, with $\zeta
_{\alpha \beta }^{i}=\zeta _{\parallel }\hat{u}_{i\alpha
}\hat{u}_{i\beta }+\zeta _{\perp }(\delta _{\alpha \beta
}-\hat{u}_{i\alpha }\hat{u}_{i\beta })$, $\zeta _{R}$ is the
rotational friction, and the mass of the rods has been set to one.
The second term on the right hand side of Eq.~(\ref{2}) describes
self propulsion as a center of mass force $F$ acting along the
long axis of each rod. This force is nonequilibrium in origin and
 arises from
an internal or external propulsion mechanism. The random forces $\bm%
\eta _{i}$ and $\eta _{i}^{R}$ describe Markovian white noise with
correlations $\left\langle \eta _{i\alpha }\left( t\right) \eta
_{j\beta }\left( t^{\prime }\right) \right\rangle =\Delta _{\alpha
\beta }^{i}\delta _{ij}\delta \left( t-t^{\prime }\right) $ and
$\left\langle \eta _{i}^{R}\left( t\right) \eta _{j}^{R}\left(
t^{\prime }\right) \right\rangle =\Delta ^{R}\delta _{ij}\delta
\left( t-t^{\prime }\right) $.  For simplicity  we assume
the equilibrium-like form 
$\Delta _{\alpha \beta }^{i}=2k_{B}T_a\zeta_{\alpha \beta }^{i}$ and 
$\Delta ^{R}=2k_{B}T_a\zeta^{R}/I$, with $I=\ell ^{2}/12$ the moment of inertia of the rod and $T_a$ an effective temperature defined by these
relationships.
Finally, the collision operator $T\left( i,j\right) $
generates the instantaneous
momentum transfer between rods at contact and is given by
%
\begin{eqnarray}
T\left( 1,2\right)
&=&\int_{s_{1},s_{2}}\int_{\hat{\mathbf{k}}}\left|
\mathbf{V}_{12}\cdot \widehat{\mathbf{k}}\right| \Theta \left( -\mathbf{V}%
_{12}\cdot \widehat{\mathbf{k}}\right)   \nonumber \\
&&\times \delta \left( \Gamma _{cont}\right) \left(
b_{12}-1\right) \;, \label{A.12.2}
\end{eqnarray}
%
where $\widehat{\mathbf{k}}$ is the unit normal
at the point of contact of the two rods directed from rod $2$ to
rod $1$, as shown in Fig.~\ref{Fig:collision}.  The  function $\Gamma _{cont}\left(
\mathbf{r}_{1},\mathbf{r}_{2},\bm\xi _{1},\bm\xi _{2}\right) $
is nonzero when two rods are at contact and
zero
otherwise. Here $\bm\xi _{i}$ is a vector from the center of mass of the $%
i$-th rod to the point of contact, $\bm\xi _{i}=s_{i}\hat{\mathbf{%
u}}_{i}\pm R\widehat{\mathbf{k}}$, where $-\ell /2\leq s_{i}\leq
\ell /2$ parametrizes the distance of points along the axis of
each rod from the
center of mass and $\int_{s_{i}}...\equiv \int_{-\ell /2}^{\ell /2}...ds_{i}$.
Also, $\mathbf{V}_{12}=\mathbf{v}_{1}-\mathbf{v}_{2}+\bm{\omega
}_{1}\times \bm{\xi }_{1}-{\bm\omega }_{2}\times \bm{\xi }_{2}$ is
the relative velocity of the two rods at the point of contact.
Finally, the operator $b_{12}$ replaces precollisional velocities
with their postcollisional values, as obtained by requiring
 energy and momentum conservation.    The
explicit calculation of the $T$ operator is given in \cite{ABMCMStatmech}.

\paragraph{Modified Smoluchowski equation.}

We are interested here in the overdamped limit, when inertial
effects are negligible and the low density dynamics is described
by a Smoluchowski equation for the the probability distribution
$c\left(x,t\right) $, with $x=(\mathbf{r},\theta)$, of rods at a point $\mathbf{r}$ oriented in the direction $%
\theta $. The derivation of the Smoluchowski equation for
self-propelled hard rods can be carried out following closely that
of thermal hard rods and is given in \cite{ABMCMStatmech}. Here,
we outline the key steps involved.

1.  First, the noise averaged statistical mechanics of a system
described by a set of coupled Langevin equations is given in terms
of the Liouville-Fokker-Planck equation governing the dynamics of
an $N$ particle distribution function \cite {ZwanzigBook}. This
can in turn be converted into a hierarchy of equations for reduced
distribution functions analogous to the BBGKY hierarchy for
Hamiltonian systems. At low density, neglecting two particle
correlations, the first equation of the hierarchy gives a closed
Boltzamnn-Fokker-Planck equation for the one particle distribution
function $f\left(x,p,t\right) $, with $p=(\mathbf{v},\omega)$.

2.  The probability distribution is $c\left(x,t\right)
=\int_{p}f(x,p,t)$. In the regime of large friction,  the
velocities of the rods decay to a stationary value on microscopic
time scales.  We use an approximate solution of the noninteracting
Fokker-Planck equation  valid in the large friction regime,
$f\left(x,p,t\right) =c\left( x ,t\right) f_{M}\left(p |\theta
\right) $,
with $f_{M}\sim\exp \left( -\frac{1}{2k_{B}T_a}\left( \mathbf{v}%
-v_{0}\widehat{\mathbf{u}}\right) ^{2}-\frac{1}{2k_{B}T_a}I\omega
^{2}\right) $ a Maxwellian distribution centered at the
self-propulsion velocity $v_{0}\hat{\mathbf{u}}$.
With this
ansatz, the Bolztmann-Fokker-Planck equation can be transformed to
a closed equation for the spatial probability distribution, $c$.

3. To obtain this closed equation we need to evaluate the mean
force and  torque on a given rod due to all other rods in the
fluid, namely $\langle T(1,2)\mathbf{v}_{1}\rangle _{M}$ and
$\langle T(1,2)\omega _{1}\rangle _{M}$, where $\left\langle
{...}\right\rangle _{M}=\int_{p_1,p_2}...
f_M(p_1|\theta_1)f_M(p_2|\theta_2)$.  In the absence of self
propulsion, this average can be readily carried out and yields the
Onsager excluded volume interaction. For finite self propulsion,
$f_{M}$ depends on the angular coordinate and hence averaging over
velocities induces
 orientational correlations that cannot be incorporated exactly.
 To make progress, we let ${\bf v}'_i={\bf v}_i-\hat{\bf u}_iv_0$ in the calculation of the velocity averages and then neglect the coupling between velocity and angular correlations by
 approximating $\langle
T(1,2)\mathbf{v}_{1}\rangle _{M}\simeq\langle
T(1,2)\mathbf{v}_{1}\rangle _{M|_{v_{0}=0}}+\langle
T(1,2)\mathbf{v}_{1}\rangle _{v_{0}}$, where the second term is
averaged over $f_{v_0}(p_1)f_{v_0}(p_2)$, with
$f_{v_0}(p_i)=\delta \left( \mathbf{v}_i-v_{0}\mathbf{\hat{u}_i}\right) \delta
\left( \omega_i \right) $.

The result is the modified Smoluchowski equation:
%
\begin{eqnarray}
\partial _{t}c+v_{0}\partial _{\parallel }c &=&D_{R}\partial _{\theta
}c+(D_{\parallel }+D_{S})\partial _{\parallel }^{2}c+D_{\perp
}\partial
_{\perp }^{2}c  \nonumber \\
&&-\frac{1}{I\zeta _{R}}\partial _{\theta }\tau _{ex}-\nabla
\cdot \bm\zeta
^{-1}\cdot \mathbf{F}_{ex}  \nonumber \\
&&-\frac{1}{I\zeta _{R}}\partial _{\theta }\tau _{SP}-\nabla \cdot
\bm\zeta ^{-1}\cdot \mathbf{F}_{SP}\;,  \label{Smol}
\end{eqnarray}
%
where $\partial _{\parallel }=\hat{\mathbf{u}}\cdot \bm\nabla $ and $\bm%
\partial _{\perp }=\bm\nabla -\hat{\mathbf{u}}(\hat{\mathbf{u}}\cdot \bm%
\nabla )$.
The convective term on the left hand side of
(\ref{Smol}) is a trivial consequence of self propulsion and
describes mass flux along the long axis of the rod. The first three
terms on the right hand side of the equation describe
translational diffusion longitudinal ($D_{\parallel }$) and
transverse ($D_{\perp }$) to the rod's long axis and rotational
diffusion ($D_{R}$). For long thin rods $D_{\Vert }=2D_{\bot }=D$. At low density
$D=k_{B}T_a/\zeta
_{\Vert }$ and $D_{R}=6D/\ell ^{2}$.  A novel consequence of self-propulsion is the
enhancement of longitudinal diffusion by $D_{S}=v_{0}^{2}/\zeta
_{\Vert }$. This can be understood by noting that a diffusing rod
performs a random walk with a step length $x_{\alpha }=\zeta
_{\alpha \beta }^{-1}v_{\beta }$. For thermal systems the rod's
velocity is isotropic on average and has magnitude $v_{th}\sim
\sqrt{k_{B}T_a}$. In this case the anisotropy of
diffusion arises solely from the anisotropy of the friction
tensor. For self-propelled rods the step length along the long
direction of the rod is enhanced, yielding an additional
contribution to the longitudinal diffusion coefficient.
Equivalently, longitudinal diffusion of a self-propelled rod
can be reformulated as a persistent random walk where the rod has a bias $%
\sim v_{0}$ towards steps along its long axis \cite{PersitentRW}.
The next three terms in (\ref{Smol}) describe excluded volume
effects within the mean-field approximation due to Onsager. The
corresponding forces and torque can be derived from the familiar
excluded volume potential as $\tau _{ex}=-\partial _{\theta
}V_{ex}$ and $\mathbf{F}_{ex}=-\bm\nabla V_{ex}$, with
$V_{ex}(x_1) =k_{B}T_ac(x_1,t)\int_{\bm\xi _{12}}\int_{\mathbf{\hat{u}}%
_{2}}\left| \mathbf{\hat{u}}_{1}\times \mathbf{\hat{u}}_{2}\right|
c\left( \mathbf{r}_{1}+{\bm\xi _{12}},\theta _{2},t\right) $, with
${\bm\xi _{12}}=\bm\xi _{1}-\bm\xi _{2}$. Finally, $\tau _{SP}$
and $\mathbf{F}_{SP}$ describe, within a mean-field approximation,
the additional torque and force due to anisotropic linear and
angular momentum transfer during the collision of two
self-propelled rods,
%
\begin{eqnarray}
\left(
\begin{array}{c}
\mathbf{F}_{SP} \\
\tau _{SP}
\end{array}
\right) &=&v_{0}^{2}\int_{x_2,s_{1},s_{2},\hat{\mathbf{k}}}
\left(\begin{array}{c}\widehat{\mathbf{k}} \\
\hat{\bf z}\cdot ( \bm\xi _{1}\times \widehat{\mathbf{k}})
\end{array}\right)
[\hat{\bf z}\cdot(\hat{\bf u}_1\times\hat{\bf u}_2) ]
^{2}\nonumber\\
&&\times\Theta( -
\mathbf{\hat{u}}_{12} \cdot
\widehat{\mathbf{k}})
\delta \left( \Gamma _{cont}\right)
c(x _{1},t)c(x _{2},t),\label{SPforces}
\end{eqnarray}
%
with $ \mathbf{\hat{u}}_{12}= \mathbf{\hat{u}}_{1}- \mathbf{\hat{u}}_{2}$.
In Onsager's mean field model,
two thin rods of length $\ell $ exchange
an average momentum $%
\left\langle |\Delta {\bf v}|\right\rangle \nu_{coll} \sim k_{B}T_a/\ell $
per unit time upon collision, with $%
\left\langle |\Delta {\bf v}|\right\rangle \sim \sqrt{k_{B}T_a}$  and $\nu_{coll}=v_{th}/\ell \sim\sqrt{k_{B}T_a}/\ell$. When
rods are self propelled there are anisotropic contributions to both
the momentum exchanged ($\left\langle |\Delta {\bf v}|\right\rangle \sim
v_{0}| \hat{\mathbf{u}}_{1}\times
\hat{\mathbf{u}}_{2}| $) and the collision rate
($\nu_{coll}\sim v_{0}|\hat{\mathbf{u}}_{1}\times \hat{%
\mathbf{u}}_{2}|/\ell$). These yield the new anisotropic steric forces and torques in Eq.~(\ref{SPforces}).

\paragraph{Hydrodynamics.}

We now use the modified Smoluchowski equation to obtain
coarse-grained equations that describe the dynamics of the systems
on wavelengths long compared to the length of the rods and on time
scales long compared to the collision time. In this regime the
dynamics is controlled by the ``slow variables'' corresponding to
the conserved densities (here only the concentration of filaments
$\rho =\int_{\widehat{\mathbf{u}}}c(x,t)$ ) and the fields
associated with possible broken symmetries.
%
%
In a liquid of self-propelled rods, both polar and nematic order
are possible, described by a polarization vector
$\mathbf{P}(\mathbf{r},t)=\int_{\widehat{\mathbf{u}}}\widehat{\mathbf{u}}c(x,t)$
and the nematic alignment tensor $Q_{\alpha \beta
}(\mathbf{r},t)=\int_{\widehat{\mathbf{u}}}(\widehat{u}_{\alpha
}\widehat{u}_{\beta }-\frac{1}{2}\delta _{\alpha \beta })c(x,t)$,
respectively. Since each rod has a self propulsion velocity
$v_{0}\widehat{\mathbf{u}}$, the polarization is also proportional
to the self propulsion flow field. The equations for these
continuum fields are obtained by taking the corresponding moments
of the Smoluchowski equation (\ref{Smol}) and are given by
\begin{eqnarray}
\partial _{t}\rho +v_{0}\nabla \cdot \mathbf{P}=D_\rho \nabla ^{2}\rho
+D_Q \bm\nabla \bm\nabla :\rho{\bf Q}  \label{2.4}
\end{eqnarray}
\begin{widetext}
\begin{eqnarray}
\partial _{t}\mathbf{P}+D_{R}\mathbf{P}-\lambda\mathbf{P}\cdot{\bf Q}+v_{0}\bm\nabla \cdot
{\bf Q}+\frac{v_{0}}{2}\bm\nabla \rho
+\lambda'[
3({\bf P}\cdot \bm\nabla) \mathbf{P}-\frac{1}{2}\bm\nabla
P^{2}-\mathbf{P}\bm\nabla \cdot \mathbf{P}]
=D_{b} \nabla ^{2}\mathbf{P}%
+(D_{spl}-D_{b})\bm\nabla \bm\nabla \cdot
\mathbf{P} \label{2.5}
\end{eqnarray}
\begin{eqnarray}
\partial _{t}{\bf Q}+4D_{R}\left( 1-\frac{\rho }{\rho _{IN}}%
\right) {\bf Q}+v_{0}{\bf F} =-%
\lambda''\left( \frac{3 }{5}\mathbf{P}\cdot
\bm\nabla {\bf Q}+\frac{1
}{48}{\bf Q}\bm\nabla \cdot
\mathbf{P+}\frac{1 }{48}{\bf G}+\frac{1 }{96%
}{\bf F}\right)  +\frac{D_Q}{4} (\bm\nabla \bm\nabla
-\frac12{\bf 1})\rho   \label{2.6}
\end{eqnarray}
\end{widetext}
 where  $F_{\alpha \beta }=\left(
\partial _{\alpha }P_{\beta }+\partial _{\beta }P_{\alpha }-\delta _{\alpha
\beta }\bm\nabla \cdot \mathbf{P}\right) $ and $G_{\alpha \beta
}=Q_{\alpha \gamma }\partial _{\gamma }P_{\beta }+Q_{\beta \gamma
}\partial _{\gamma }P_{\alpha }-\delta _{\alpha \beta }Q_{\sigma
\gamma }\partial _{\gamma }P_{\sigma }$. All $\lambda$ parameters
in Eqs.~(\ref{2.5}) and (\ref{2.6}) are proportional to $v_0^2$
and vanish in the absence of self propulsion. All diffusion
constants are enhanced by self-propulsion via additive terms
proportional to $D_S$. Finally,
 we have suppressed in Eqs.~(\ref{2.4}-\ref{2.6}) excluded volume corrections to
the diffusive terms, nonlinear terms of second order
in gradients, and corrections to the convective terms beyond linear in $v_0$.
The complete hydrodynamic equations with explicit expressions for the various coefficients
can be found in Ref.
\cite{ABMCMStatmech}.

The stable homogeneous stationary solution of Eqs.~(\ref{2.4}-\ref{2.6})
are the  bulk states of the self-propelled
system. Two such states are possible:
an isotropic state, with $\rho =\mathrm{constant}$,
$\mathbf{P}=0$, $Q_{\alpha \beta }=0$, and a
nematic state, with  $\rho =\mathrm{constant}$, $\mathbf{P}=0$ and $%
Q_{\alpha \beta }\not=0$. Hard core interactions and
self-propulsion modeled simply as a body force are not sufficient
to generate a bulk polar state, with $\mathbf{P}\not=0$. Either
shape or mass distribution asymmetry of the driven particles or
hydrodynamic interactions are essential to obtain a macroscopic
polar (moving) state. Self-propulsion has, however, a profound
effect on the isotropic-nematic transition which occurs at the
density $\rho _{IN}(v_{0})=\rho
_{N}/(1+\frac{v_{0}^{2}}{5k_{B}T})$, where $\rho _{N}=3/(\pi \ell
^{2})$ is the Onsager transition density. The transition occurs
where the coefficient of the term linear in $Q_{\alpha \beta }$ on
the right hand side of Eq.~(\ref{2.6}) changes sign, signaling the
unstable growth of nematic fluctuations. This enhancement of
orientational order  has been observed  in numerical simulations
of actin motor assays, where actin filaments move on a substrate
grafted with  motor proteins \cite{Lipowsky}. It arises from the
additional torque $\tau _{SP}$ that self-propelled rods experience
upon collision as compared to thermal rods. This enhances entropic
ordering
 and  aligns the rods~\cite{SPtransverse}.

\begin{figure}[tbp]
\centerline{\epsfxsize=5cm \epsfbox{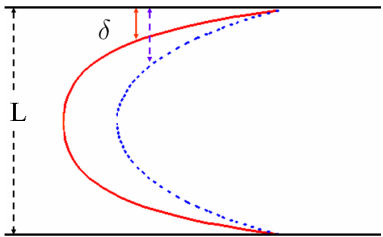}} 
\caption{ (color online) The polarization in a channel of width
$L$ for $\delta/L=0.2$ (solid) and $\delta/L=0.6$ (dashed).}
\label{Fig:Polarization}
\end{figure}

Although no bulk polar order is possible in our system,
self-propulsion greatly enhances the length scale over which
polarization fluctuations decay. As a result boundaries pay a
crucial role in self propelled systems. To illustrate this we
consider a self-propelled 2d hard rod fluid confined in the
channel of width $L$ between two boundaries, as shown in Fig.~\ref{Fig:Polarization}. We
assume that the boundaries induce polarity by forcing all
rods to align in the same direction, i.e.,  $%
P_{x}(-L/2)=P_{x}(L/2)=P_{0}$. In this geometry the density is
constant. One can easily solve for the polarization profile across
the channel with the
result $P_{x}(y)=P_{0}\cosh (y/\delta )/\cosh(L/2\delta)$, where $\delta =\sqrt{D_{b}/D_{R}}%
=\ell /2\sqrt{5/2+v_{0}^{2}/k_{B}T}$ is the boundary layer width
over which the polarization penetrates in the channel. In the
absence of self-propulsion $\delta \sim \ell $, i.e.,  a finite
polarization at the boundary decays (via rotational diffusion)
over a length scale of  order $\ell$. For
large self-propulsion velocity, $\delta \sim |v_{0}|$. If $L\sim
\delta $ the entire channel is effectively polarized. We stress
that numerical simulations of self-propelled rods on a substrate
have indeed observed large correlated regions of finite
polarization, but never an ordered bulk state. We expect that the
boundary layer length $\delta $ also sets the scale of
correlations in bulk systems. Finally, as shown in \cite{Aparna2},
Eqs.~(\ref{2.4}-\ref{2.6}) yield  two important properties of fluctuations in
self-propelled systems. First, the isotropic state can support
sound-like propagating density waves for a range of wavevectors
above a critical value of $v_0$. Secondly, large number
fluctuations always destabilize the homogeneous nematic state. We
refer the reader to Ref. \cite{Aparna2} for a complete description
of both results.

In summary, we have analyzed a simple model
that captures two crucial properties of self-propelled systems:  the orientable shape
of the particles and the self propulsion.
Using the tools of nonequilibrium statistical mechanics
we have derived a modified Smoluchowski equation for SPP
and used it to identify the microscopic origin of several observed
or observable large scale phenomena.

This work was supported by the NSF on grants DMR-0305407 and
DMR-0705105.


\begin{thebibliography}{99}
\bibitem{SPtransverse}
No enhancement of orientational
order occurs if the self-propulsion velocity is normal to
the rods' long axis.
\end{thebibliography}
\end{document}